# Amplification without inversion, fast light and optical bistability in a duplicated two-level system

Lida Ebrahimi Zohravi, Azar Vafavard, Mohammad Mahmoudi

Department of Physics, University of Zanjan, University Blvd, 45371-38791, Zanjan, Iran

Tel.: +98-241-5152521

Fax: +98-241-5152264

E-mail: mahmoudi@znu.ac.ir

**Abstract.** The optical properties of a weak probe field in a duplicated two-level system are investigated in multi-photon resonance (MPR) condition and beyond it. It is shown that, by changing the relative phase of applied fields, the absorption switches to the amplification without inversion in MPR condition. By applying the Floquet decomposition to the equation of motion beyond MPR condition, it is shown that the phase-dependent behavior is valid only in MPR condition. Moreover, it is demonstrated that the group velocity of light pulse can be controlled by the intensity of applied fields and the gain-assisted superluminal light propagation (fast light) is obtained in this system. In addition, the optical bistability (OB) behavior of the system is studied beyond MPR condition. We apply an indirect incoherent pumping field to the system and it is found that the group velocity and OB behavior of the system can be controlled by incoherent pumping rate.

## 1. Introduction

The study of the optical response of a medium is one of the most important fields in optics. Quantum coherence has a major role in determination of the optical properties of the systems and has various applications in optical physics [1]. It has been used to establish the Lasing without inversion (LWI) [2], electromagnetically induced transparency (EIT) [3], refractive index enhancement [4], optical bistability [5], slow [6] and fast light [7]. In superluminal light propagation the group velocity of light pulse in a dispersive medium can exceed the speed of light in vacuum and even can be negative [8]. However, it does not violate Einstein's principle of special relativity. It is generally believed that no complete information can be sent faster than light speed in vacuum [9]. Recently the duplicated two-level (DTL) system has been extensively studied, because of its interesting feature and potentiality of applications. The interaction of such system with femtosecond coherent pulses has been studied by Delagnes and Bouchene and the efficient modulation of the medium gain for the probe field was demonstrated due to the interference [10]. Slow light propagation through DTL system has been studied and a new method based on Zeeman coherence oscillation for slowing light has been presented in an electromagnetically induced transparency window [11]. Coherent control of the effective susceptibility through wave mixing in this system has been studied in MPR condition and it was presented a phase-dependent behaviour due to the different quantum path associated with conjugate phase [12]. The spatial interference of resonance florescence from two DTL atoms driven by two orthogonally polarized fields has been investigated in MPR condition. It was shown that the interference pattern can be recovered in the fluorescence light of strongly driven atoms due to the effect of the relative phase of applied fields on the populations and atomic coherences [13]. Controllable OB and multistability of a DTL atomic system has been studied and phase-dependent behaviour has been obtained in MPR condition [14]. Recently the phase-control of the Goos-Hänchen shift using a DTL atomic medium has been reported in MPR condition [15].

It is well known that the optical properties of a closed-loop atomic system, in MPR condition, depends on the relative phase of applied fields [16]. This is due to the wave mixing of applied fields which is not generally allowed beyond MPR condition [17].

In this manuscript, we investigate the optical properties of a DTL atomic system in MPR condition and beyond it. We obtain the amplification without inversion (AWI) in MPR condition and control it by relative phase of applied field via the phase conjugation effect. Beyond MPR condition, the Floquet decomposition is applied to solve the time-dependent differential equations of motion. We find that in general, the optical properties of the system are not phase-dependent. Moreover, the gain-assisted superluminal light propagation is obtained in this system. Finally the OB behaviour of the system is calculated beyond MPR condition.

By applying an indirect incoherent pumping field to the system, it is demonstrated that the group velocity of light pulse as well as the OB behaviour can be controlled by intensity of either coherent or incoherent pumping fields.

## 2. Theoretical Analysis

### 2.1. Model and equations

We consider a DTL system with two degenerate ground states $|1\rangle$ and $|2\rangle$ and two degenerate excited states $|3\rangle$ and $|4\rangle$. The transitions $|1\rangle - |3\rangle$ and $|2\rangle - |4\rangle$ with identical $m_F$ (with central resonance frequency $\omega_{31} = \omega_{24}$) are excited by a $\pi$-polarized control field $\vec{E}_c(t,y) = E_c \hat{e}_z \exp[-i(\omega_c t - k_c y)] + c.c.$ with Rabi frequency $\Omega_c$. A $\sigma$-polarized probe field $\vec{E}_p(t,y) = E_p \hat{e}_x \exp[-i(\omega_p t - k_p y)] + c.c.$ with Rabi frequency $\Omega_p$ is applied to the transitions $|1\rangle - |4\rangle$ and $|2\rangle - |3\rangle$ with different $m_F$ as shown in figure 1(a). As a realistic example, we consider $F_{1/2} \to F_{1/2}$ transition (e.g., $^2S_{1/2}F_{1/2} \to {}^2P_{1/2}F_{1/2}$ transition of $^6Li$ at $671 nm$) excited by two copropagating, linearly polarized laser fields.

The density matrix equations of motion in the rotating wave approximation and in the rotating frame are given by

$$i\dot{\rho}_{11} = [(\Omega_c \rho_{13} + \Omega_p \rho_{14} e^{-i\Phi(t,\vec{r})}) - cc] + i\Gamma(\rho_{33} + 2\rho_{44})/3,$$

$$i\dot{\rho}_{22} = [(-\Omega_c \rho_{24} + \Omega_p \rho_{23} e^{-i\Phi(t,\vec{r})}) - cc] + i\Gamma(\rho_{44} + 2\rho_{33})/3,$$

$$i\dot{\rho}_{33} = [-(\Omega_c \rho_{13} + \Omega_p \rho_{23} e^{-i\Phi(t,\vec{r})}) - cc] - i\Gamma\rho_{33},$$

$$i\dot{\rho}_{44} = [(\Omega_c \rho_{24} - \Omega_p \rho_{14} e^{-i\Phi(t,\vec{r})}) - cc] - i\Gamma\rho_{44},$$

$$i\dot{\rho}_{31} = \Omega_c(\rho_{33} - \rho_{11}) + \Omega_p e^{-i\Phi(t,\vec{r})}(\rho_{34} - \rho_{21}) + \overline{\Delta}_c^* \rho_{31},$$

$$i\dot{\rho}_{42} = \Omega_c(\rho_{22} - \rho_{44}) + \Omega_p e^{-i\Phi(t,\vec{r})}(\rho_{43} - \rho_{12}) + \overline{\Delta}_c^* \rho_{42},$$

$$i\dot{\rho}_{12} = -(\Omega_c \rho_{14} + \Omega_c^* \rho_{32}) + (\Omega_p e^{-i\Phi(t,\vec{r})}\rho_{13} - \Omega_p^* e^{i\Phi(t,\vec{r})}\rho_{42}) - i\Gamma_{zg}\rho_{12},$$

$$i\dot{\rho}_{34} = -(\Omega_c \rho_{14} + \Omega_c^* \rho_{32}) + (-\Omega_p e^{-i\Phi(t,\vec{r})}\rho_{24} + \Omega_p^* e^{i\Phi(t,\vec{r})}\rho_{31}) - i\Gamma_{ze}\rho_{34},$$

$$i\dot{\rho}_{41} = \Omega_c(\rho_{21} + \rho_{43}) + \Omega_p e^{-i\Phi(t,\vec{r})}(\rho_{44} - \rho_{11}) + \overline{\Delta}_c^* \rho_{41},$$

$$i\dot{\rho}_{32} = -\Omega_c(\rho_{12} + \rho_{34}) + \Omega_p e^{-i\Phi(t,\vec{r})}(\rho_{33} - \rho_{22}) + \overline{\Delta}_c^* \rho_{32}, \tag{1}$$

where $\overline{\Delta}_c = \Delta_c + i\Gamma_d$, $\Delta_c = \omega_{31} - \omega_c$, $\Phi(t,\vec{r}) = \Delta t - (k_p \hat{e}_y - \vec{k}_c).\vec{r} + \phi$ and $\Delta = \omega_p - \omega_c$. The parameter $\phi$ is the initial relative phase of applied fields. All the coherence except the ones between Zeeman levels relax with the rate $\Gamma_d$. In the absence of non-radiative dephasing processes, $\Gamma_d$ reduces to $\Gamma/2$. The excited state Zeeman coherence $\rho_{34}$ and the ground Zeeman coherence $\Gamma_{zg}$ relax with the rate $\Gamma_{ze}$ and $\Gamma_{zg}$, respectively. In pure radiative dephasing $(\Gamma_{ze}, \Gamma_{zg})$ reduces to $(\Gamma, 0)$ [12]. The decay rates from excited states to the ground states are assumed $\Gamma/3$ for identical $m_F$ and $2\Gamma/3$ for different $m_F$. The equations (1) can be simplified, due to the symmetry of the system, by a suitable change of variable. We define new variables as

$$\rho_p = \rho_{32} + \rho_{41}, \rho_c = \rho_{31} - \rho_{42}, n_g = \rho_{11} + \rho_{22}, n_e = \rho_{33} + \rho_{44}, \rho_{zg} = \rho_{12} - \rho_{21}, \rho_{ze} = \rho_{34} - \rho_{43}. \tag{2}$$

Then equations (1) reduce to

$$i\dot{n}_g = (\Omega_c \rho_c^* + \Omega_p e^{-i\Phi(t,\vec{r})} \rho_p^* - cc) + i\Gamma(1 - n_g) - iR\, n_g,$$

$$i\dot{\rho}_c = \Omega_c (n_e - n_g) + \Omega_p e^{-i\Phi(t,\vec{r})} (\rho_{zg} + \rho_{ze}) + \overline{\Delta}_c^* \rho_c - iR\, \rho_c,$$

$$i\dot{\rho}_p = -\Omega_c (\rho_{zg} + \rho_{ze}) + \Omega_p e^{-i\Phi(t,\vec{r})} (n_e - n_g) + \overline{\Delta}_c^* \rho_p - iR\, \rho_p,$$

$$i\dot{\rho}_{zg} = (-\Omega_c \rho_p^* + \Omega_p e^{-i\Phi(t,\vec{r})} \rho_c^* + cc) - i\Gamma_{zg}\, \rho_{zg},$$

$$i\dot{\rho}_{ze} = (-\Omega_c \rho_p^* + \Omega_p e^{-i\Phi(t,\vec{r})} \rho_c^* + cc) - i\Gamma_{ze}\, \rho_{ze}. \tag{3}$$

Parameter $R$ denotes the indirect incoherent pumping rate which is applied to the probe transitions.

## 2.2. Linear susceptibility and group velocity

The response of the system to the probe fields is determined by the susceptibility $\chi = \chi' + i\chi''$, which is defined as [18]:

$$\chi(\omega_p) = \frac{2\alpha_0 \Gamma_d}{k_p} \frac{\rho_p(\omega_p)}{\Omega_p e^{-i\phi}} \tag{4}$$

where $\alpha_0 = N D^2 \omega_p /(2c\hbar\varepsilon_0 \Gamma_d)$ and $k_p = \omega_p / c$. $N$ is the atom number density in the medium and $D$ denotes the dipole moment of transitions. The real and imaginary parts of $\chi$ correspond to the dispersion and the absorption of the probe field, respectively. The transition rate and dipole moment of the probe transition can be considered as $\Gamma = 37\,MHz$ and $D = 2.81\times 10^{-29}\,C.m$, respectively. Then the atom density $N = 4.36\times 10^{11}\,atom/cm^3$ and probe Rabi frequency $|\Omega_p| = 0.01\Gamma$, lead to $2\alpha_0 \Gamma_d / k_p \Omega_p \cong 1$.

The group velocity of light pulse is determined by the slope of dispersion. We introduce the group index $n_g = \dfrac{c}{v_g}$, where the group velocity $v_g$ is given by

$$v_g = \frac{c}{1 + \dfrac{1}{2}\chi'(\omega_p) + \dfrac{\omega_p}{2}\dfrac{\partial \chi'(\omega_p)}{\partial \omega_p}} = \frac{c}{n_g}. \tag{5}$$

In a dispersive medium, the refractive index depends on the frequency and then the different frequency components of a light pulse have different phase velocities. Therefore the group velocity of light pulse in such a medium can exceed the velocity of light in vacuum, leading to the superluminal light propagation. In our notation the negative slope of dispersion corresponds to the anomalous dispersion, while the positive slope shows the normal dispersion. Moreover, the positive (negative) value in imaginary part of susceptibility shows the absorption (gain) for the probe field.

*2.3. Optical bistability*

OB has been extensively studied because of its wide applications in optical transistors, memory elements and all optical switching [5]. The optical bistability for a Bose-Einstein condensate of atoms in a driven optical cavity with a Kerr medium has been studied and it was found that both the threshold point of optical bistability transition and the width of optical bistability hysteresis can be controlled by appropriately adjusting the Kerr interaction between the photons [19]. Recently we have studied the OB behavior in a two-level atom [20], as well as three-level atoms [21], beyond two-photon resonance condition. In this paper, we study the OB behavior of the DTL system, beyond MPR condition. Now, we put the DTL systems in a unidirectional ring cavity as shown in Fig.1-b. For simplicity, we assume that mirrors 3 and 4 have 100% reflectivity, and the intensity reflection and transmission coefficient of mirrors 1 and 2 are $\bar{R}$ and $\bar{T}$ (with $\bar{R}+\bar{T}=1$). We consider a collection of $N$ DTL atomic systems contained in a cell of length $L$. The total electromagnetic field can be written as $\vec{E}=\vec{E}_c(t,y)+\vec{E}_p(t,y)+c.c.$, where only the probe field circulates in the ring cavity. Then under slowly varying envelope approximation, the dynamic response of the probe field is governed by Maxwell's equation

$$\frac{\partial E_p}{\partial t}+c\frac{\partial E_p}{\partial z}=i\frac{\omega_p}{2\varepsilon_0}P(\omega_p) \quad (6)$$

where $\varepsilon_0$ is the permittivity of free space. $P(\omega_p)$ is the induced polarization in the probe transitions and is given by $P(\omega_p)=ND\rho_p$. The boundary conditions between the incident field $E_p^I$ and the transmitted field $E_p^T$ for a perfectly tuned ring cavity lead to

$$E_p(L)=E_p^T/\sqrt{\bar{T}}$$

$$E_p(0)=\sqrt{\bar{T}}E_p^I+\bar{R}E_p(L). \quad (7)$$

The feedback mechanism of the probe field inside the ring cavity can induce the OB behavior. In the mean-field limit by using the boundary conditions, i.e., equations (7), the input-output relation is given by

$$y=2x-iC\rho_p, \quad (8)$$

where $C=N\omega_p LD^2/\hbar\varepsilon_0 c\bar{T}$ is the usual cooperation parameter. Parameters $x=DE_p^T/\hbar\sqrt{\bar{T}}$ and $y=DE_p^I/\hbar\sqrt{\bar{T}}$ are the normalized output and input fields, respectively.

## 2.4. MPR condition

We assume that the MPR condition ($\Delta = 0$, $k_p \hat{e}_y = \vec{k}_c$) is fulfilled by the applied fields. Then equations (3) do not have the explicitly time-dependent argument and system has a steady state solution. Under the weak probe field approximation and for $\Delta_c = 0$, $R = 0$, the following simple analytical expressions for the total population of Zeeman ground states and probe susceptibility are obtained

$$n_g = \frac{1}{2}\left[\frac{|\Omega_c|^2 + 0.5\Gamma\Gamma_d}{|\Omega_c|^2 + 0.25\Gamma\Gamma_d}\right], \tag{9}$$

$$\chi(\omega_p) \propto \frac{\rho_p}{\Omega_p e^{-i\phi}} = \frac{i\Gamma_d\,\Gamma_{zg}\Gamma_{ze}\Gamma}{A} + \frac{i\Gamma\Gamma_{zg}|\Omega_c|^2}{A} + \frac{\Omega_p^*}{\Omega_p}e^{2i\phi}\left[\frac{i\Gamma(2\Gamma_{ze} + \Gamma_{zg})\Omega_c^2}{A}\right], \tag{10}$$

$$A = (\Gamma\Gamma_d + 4|\Omega_c|^2)\left[\Gamma_d\,\Gamma_{zg}\Gamma_{ze} + 2(\Gamma_{ze} + \Gamma_{zg})|\Omega_c|^2\right].$$

Equation (9) shows that the population of Zeeman ground states does not depend on the relative phase of applied fields. Moreover $n_g$ is more than the half of the total population and then the population inversion cannot establish in this system, even by changing the amplitude and relative phase of applied fields.

But the situation is completely different in equation (10). Three terms involve different physical processes which have simple interpretations. The first term shows the direct response of the medium to the probe field. The second term indicates the cross Kerr effect. In the absence of ground Zeeman coherence relaxation, these two contributions are ignored [12]. The third term is the phase conjugate effect which determines the phase-dependent response of the medium to the probe field. Note that the direct response of the medium to the probe field as well as the cross Kerr nonlinearity effect are not phase-dependent, however in MPR condition, the contributions of all three terms of equation (10) are mixed together and the optical behavior becomes phase-dependent.

## 2.5. Beyond MPR condition

Beyond MPR condition, the explicitly time-dependent terms in equations (3) do not permit to obtain the constant steady state and the Floquet decomposition should be applied to the set of equations (3) [22-23]. The elements of density matrix can be expanded as

$$\rho_{ij} = \rho_{ij}^{(0)} + \sum_{m=1}(\rho_{ij}^{(m)}e^{-im\Phi(r,t)} + \rho_{ij}^{(-m)}e^{im\Phi(r,t)}) \tag{11}$$

where $m$ is an integer and the maximum $m$ indicate order of approximation. In the presence of incoherent pumping, the zero order stationary state solutions are given by

$$n_e^{(0)} = \frac{2|\Omega_c|^2(\Gamma_d + R) + R|\overline{\Delta}_c|^2}{4|\Omega_c|^2(\Gamma_d + R) + (\Gamma + R)|\overline{\Delta}_c|^2}, \quad (12\text{-a})$$

$$\rho_p^{(0)} = \rho_{zg}^{(0)} = \rho_{ze}^{(0)}. \quad (12\text{-b})$$

Using equation (12-a) the inverse of $n_e^{(0)}$ can be obtained as

$$\frac{1}{n_e^{(0)}} = 2\left(1 + \frac{(\Gamma - R)}{2}\frac{|\overline{\Delta}_c|^2}{2|\Omega_c|^2(\Gamma_d + R) + R|\overline{\Delta}_c|^2}\right). \quad (13)$$

According to the equation (13), it is clear that for $R < \Gamma$ ($R > \Gamma$) the total population of excited states is less (more) than of lower levels and then system shows the absorption (gain). The saturation effect is also obtained for $R = \Gamma$.

The solution for the probe susceptibility and group index in first order, for $\Omega_c, \Delta_c \ll \Gamma_d$ can be written as

$$\chi(\omega_p) = \frac{2\alpha_0\Gamma_d}{k_p}\Delta(\Gamma_d + R)\left(\frac{\Gamma - R}{\Gamma + R}\right)\frac{2|\Omega_c|^2 - \Delta^2 + i\Delta(\Gamma_d + R)}{4|\Omega_c|^4 + (\Gamma_d + R)^2\Delta^2}, \quad (14)$$

$$n_g = \frac{c}{v_g} - 1 = \frac{\omega_p\alpha_0\Gamma_d(\Gamma_d + R)}{k_p}\left(\frac{\Gamma - R}{\Gamma + R}\right)\frac{|\Omega_c|^2[4|\Omega_c|^4 - \Delta^2(\Gamma_d + R)^2]}{[4|\Omega_c|^4 + (\Gamma_d + R)^2\Delta^2]^2}. \quad (15)$$

Note that the stationary solutions equations (12)-(15) do not depend on the relative phase of applied fields. Our analytical results for $R = 0$ are in good agreement with the results of reference [24].

The coefficient $(\Gamma - R)/(\Gamma + R)$ has a major role in determination of subluminal or superluminal light propagation. For $\Gamma > R$ the group index is positive, corresponding to the subluminal condition. By increasing the incoherent pumping rate to $R > \Gamma$, the group index becomes negative and the pulse light propagates in superluminal regime. For $R = \Gamma$, system shows the saturation behavior and then the group velocity is equal to the speed of light in vacuum. Then by applying the incoherent pumping field to the system, we introduce an additional parameter for controlling the group velocity and switching from subluminal to superluminal light propagation.

## 3. Results and discussion

## 3.1. AWI in the absence of the incoherent pumping

Now, we are interested in summarize the numerical results of equations (3), in MPR condition. We assume $\Gamma=1$ and all parameters are scaled by $\Gamma$. Firstly, we investigate the phase-dependent optical properties of the system through the numerical calculations in the absence of incoherent pumping. Figure 2 shows the absorption of probe field (a) and population of Zeeman ground states (b) versus control field detuning for different values of relative phases of applied fields. Using parameters $\Gamma=1$, $\Gamma_d=0.5\Gamma$, $\Delta_c=0.0$, $\Omega_p=0.01\Gamma$ and $\Omega_c=0.6\Gamma$, $\phi=0$ (solid), $\pi/2$ (dashed). By changing the relative phase from $\phi=0$ to $\phi=\pi/2$, the absorption peak changes to the gain, without changing in the population of ground Zeeman states which can be explained by equations (9) and (10). The imaginary part of the susceptibility depends on $\phi$ through the coefficient $cos(2\phi)$. Then the absorption of probe field can be changed to the AWI just by changing the relative phase of applied fields.

AWI is obtained even beyond MPR condition. However it is controlled by the intensity of control field. Figure 3 shows the absorption (a) and the population of Zeeman ground states (b) versus the Rabi frequency of control field. It is shown that the population inversion does not establish for different values of the intensity, but by increasing the intensity of control field the absorption switches to the AWI. Then beyond MPR condition, the AWI can be controlled by the intensity of control field.

## 3.2. Fast light beyond MPR condition

In the following we investigate the optical properties of the system beyond MPR condition. Figure 4 shows the dispersion (solid) and the absorption (dashed) of probe field versus $\Delta/\Gamma$ when the incoherent pumping field is switched off. Using parameters are $\Gamma=1$, $\Gamma_d=0.5\Gamma$, $\Delta_c=0.0$, $\Omega_p=0.01\Gamma$ and $\Omega_c=0.1\Gamma(a), 0.6\Gamma(b)$. Note that we choose just the contribution of the probe field ($\rho_p^{(1)}$) in the susceptibility. An investigation on figure 4 shows that for small values of $\Omega_c$ the slope of dispersion around zero probe detuning is positive, accompanied by a doublet absorption in the spectrum. By increasing the Rabi frequency of control field, the slope of dispersion switches from positive to negative corresponding to the superluminal light propagation. Moreover, the doublet absorption switches to the doublet gain. The related group index is plotted in figure 5. The solid (dashed) line shows the group index for $\Omega_c=0.1\Gamma(0.6\Gamma)$. The group index for $\Omega_c=0.1\Gamma$ is positive around zero probe detuning, while it is negative for $\Omega_c=0.6\Gamma$.

The other controlling parameter is the incoherent pumping field rate. We are interested in studying the effect of incoherent pumping field on the optical properties of the system. In figure 6, we present the dispersion (solid) and absorption (dashed) for $R=0.5\Gamma(a), 1.5\Gamma(b)$. Other using parameters are same as in figure 4(a). It is clear that by increasing the incoherent pumping rate ($R > \Gamma$), the slope of dispersion switches from normal to anomalous dispersion. The slope of dispersion around zero probe detuning, in the presence of incoherent pumping field, is much steeper in comparing to the results of figure 4. The related the group index versus the probe detuning in the presence of incoherent field is shown in figure 7 for $R = 0.5\Gamma(solid), 1.5\Gamma(dashed)$. The subluminal light propagation in the electromagnetically induced transparency window changes to the gain-assisted superluminal light propagation just by increasing the incoherent pumping rate.

Note that for purely quantum mechanical reasons, any gain or loss process will add noises to a transmitted light field [25]. Then the propagation of light pulse in the absence of absorption or gain is an ideal condition for noise-free propagating light pulse through a dispersive medium.

### *3.3. OB beyond MPR condition*

We are going to present the OB behavior of the system, beyond MPR condition. We include the contributions of higher order oscillation terms (up to $m = 6$) of equation (11) and plot the input-output curve in figure 8 in the absence (solid) or presence (dashed and dotted) of incoherent pumping field for $\Omega_c = 0.1\Gamma(a,b), 2\Gamma(c,d)$. Using parameters are $\Delta_c = 0.0$, $\Delta = 2.0\Gamma$, $R = 0.0 (Solid), 1\Gamma(dashed), 1.5\Gamma(dotted)$. Other using parameters are same as in figure 4(a). In parts (a) and (c), we use all of contributions of equation (11) to calculate the OB diagram, while in parts (b) and (d), we use only the contribution of the probe field. Our result does not show the difference between two parts in weak control field approximation. This is due to the fact that the control field does not scatter into the probe field frequency without using the probe transition and then ($\rho_p^{(0)} = 0$). By increasing the Rabi frequency of control field, slightly difference is appeared between the OB due to the probe contribution and due to the all of contributions. It is worth to note that for small values of incoherent pumping rate, the OB behavior is due to the absorption. For $R = \Gamma$, the system does not show the bistability behavior because of the saturation effect. By increasing the pumping rate to $R = 1.5\Gamma$ the population inversion is established and the system again shows the bistability behavior because of its gain. The gain property of the system induces a zero value for upper return point in OB behavior of the system. In the absence of incoherent pumping field and in MPR condition, our OB results confirm the results of reference [14].

### **4. Conclusion**

We studied the optical properties of DTL system in MPR condition and beyond it. It was obtained that the phase-dependent AWI can be established in MPR condition. In addition, it was demonstrated that the slope of dispersion can be controlled by intensity of the applied fields. By using the Floquet decomposition, beyond MPR condition, it was shown that absorption without inversion switches to the AWI, just by increasing the intensity of control field. Moreover, we found that the subluminal light propagation switches to the absorption-free fast light. Finally, we applied an indirect incoherent pumping field to the system and we obtained that the slope of dispersion becomes much steeper and the group velocity of the probe field switches from the subluminal to the absorption-free superluminal light propagation. Moreover, the OB behavior was investigated beyond MPR condition and the effect of incoherent pumping field on the OB behavior was studied. It was demonstrated that the incoherent pumping rate has a major role in controlling the OB behavior of the system.

**Figure Captions:**

**Figure 1.** (a) Schematic diagram of DTL system. (b) Unidirectional ring cavity with DTL sample of length $L$. $E_p^I$ and $E_p^T$ are the incident and transmitted fields, while $\vec{E}_p$ and $\vec{E}_c$ are the probe and control fields, respectively. For mirror 1 and 2 it is assumed that $\bar{R} + \bar{T} = 1$ and mirrors 3 and 4 have perfect reflectivity.

**Figure 2.** Absorption of probe field (a) and population of Zeeman ground states (b) versus control field detuning for different values of relative phases of applied fields in MPR condition. Using parameters $\Gamma = 1$, $\Gamma_d = 0.5\Gamma$, $\Delta_c = 0.0$, $\Omega_p = 0.01\Gamma$ and $\Omega_c = 0.1\Gamma$, $\phi = 0$ (solid), $\pi/2$ (dashed).

**Figure 3.** Absorption (a) and the population of Zeeman ground states (b) versus the Rabi frequency of control field for $\Delta = 0.9\Gamma$. Other using parameters are same as in figure 2.

**Figure 4.** Dispersion (solid) and absorption (dashed) of the probe field versus $\Delta/\Gamma$ when the incoherent pumping field is switched off. Using parameters are $\Gamma = 1$, $\Gamma_d = 0.5\Gamma$, $\Delta_c = 0.0$, $\Omega_p = 0.01\Gamma$ and $\Omega_c = 0.1\Gamma(a), 0.6\Gamma(b)$.

**Figure 5.** The group index versus $\Delta/\Gamma$ in the absence of incoherent pumping field. The solid (dashed) line shows the group index for $\Omega_c = 0.1\Gamma(0.6\Gamma)$. Other parameters are same as in figure 4.

**Figure 6.** Dispersion (solid) and absorption (dashed) of probe field versus $\Delta/\Gamma$ when the incoherent pumping field is switched on. It is assumed that $R = 0.5\Gamma(a), 1.5\Gamma(b)$. Other using parameters are same as in figure 4(a).

**Figure 7.** The group index versus $\Delta/\Gamma$ in the presence of incoherent pumping field. The solid (dashed) line shows the group index for $R = 0.5\Gamma(1.5\Gamma)$. Other parameters are same as in figure 4(a).

**Figure 8.** OB in the absence (solid) and in the presence (dashed and dotted) of incoherent pumping field for $\Omega_c = 0.1\Gamma(a,b), 2\Gamma(c,d)$. In parts (a) and (c), we use all of contributions of equation (10) to calculatethe OB diagram, while in parts (b) and (d), we choose just the contribution of probe field. Using parameters are $\Delta_c = 0.0$, $\Delta = 2.0\Gamma$, $R = 0.0 (Solid), \Gamma(dashed), 1.5\Gamma(dotted)$. Other using parameters are same as in figure 4(a).

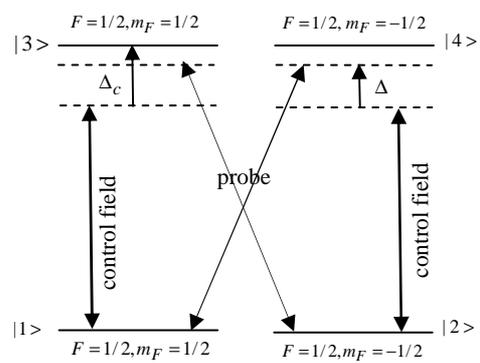

(a)

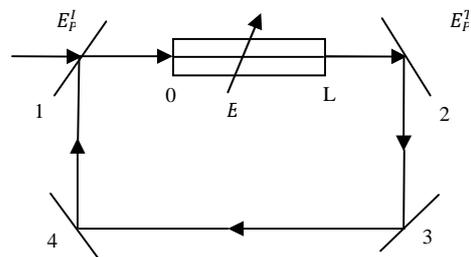

(b)

Figure 1

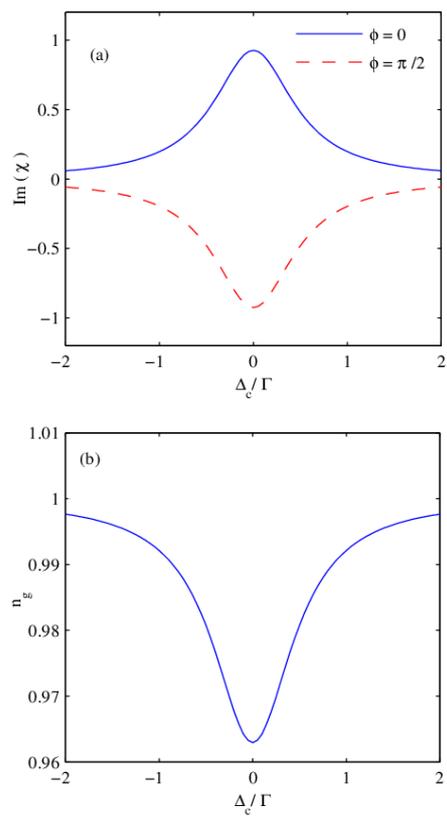

Figure 2

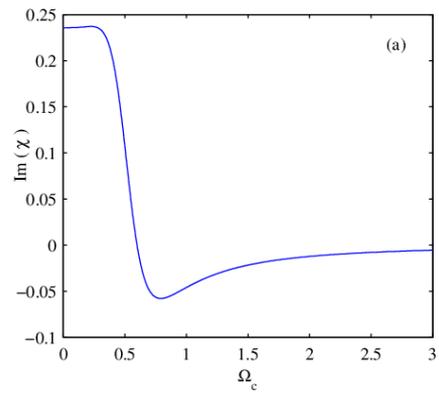

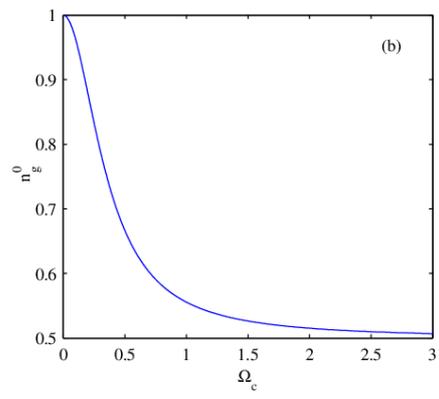

Figure 3

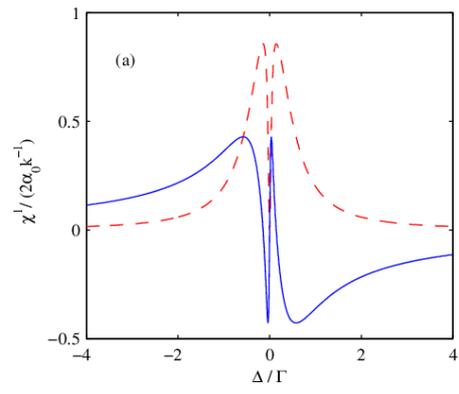

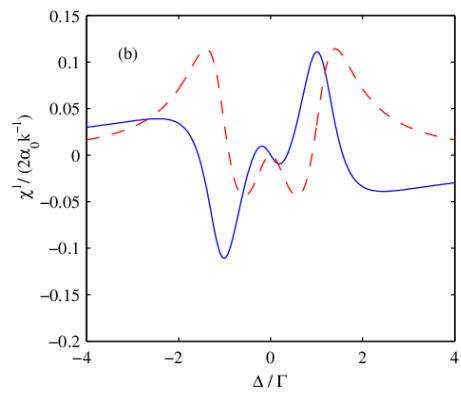

Figure 4

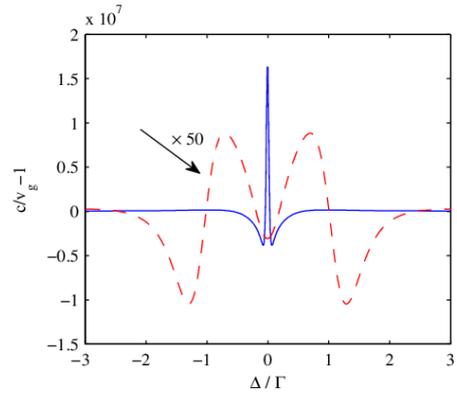

Figure 5

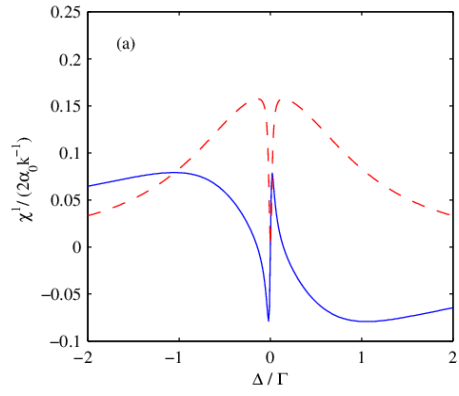

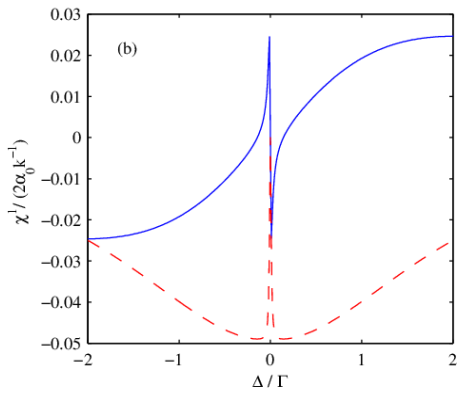

Figure 6

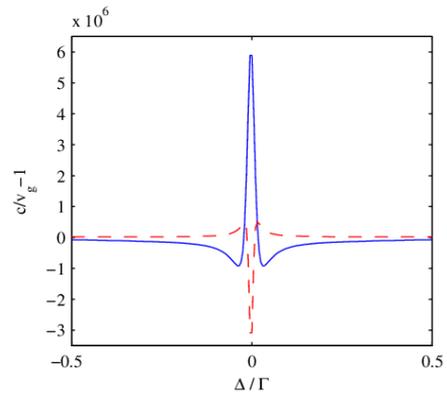

Figure 7

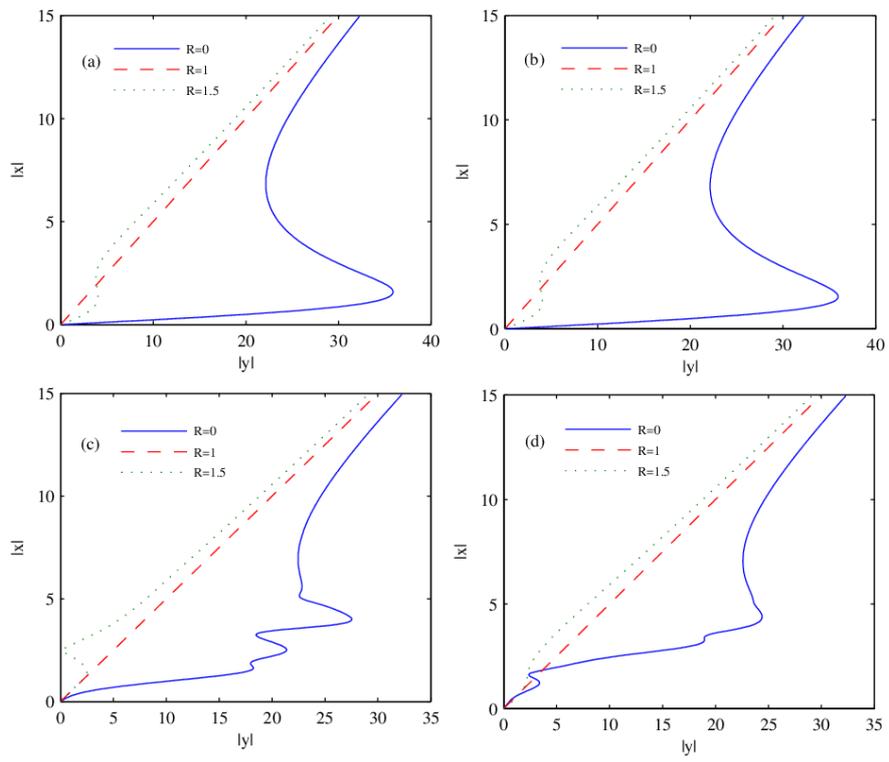

Figure 8